%% file: ms.tex
\documentclass[apj]{emulateapj}
\usepackage{amsmath}
\input{Definitions.tex}
\newcommand{\Xesc}{X_{\text{esc}}}
\newcommand{\Epar}{\vep}
\newcommand{\Ep}{\Epar_p}
\newcommand{\Enu}{\Epar_{\nu}}
\newcommand{\QUp}{Q}
\newcommand{\QMWp}{Q_{\rm MW}}
\newcommand{\Qunit}{\erg\Mpc^{-3}\yr^{-1}}

\begin{document}
\newcommand{\vv}{\textrm{v}}
\title{The energy production rate density of cosmic rays in the local universe is $\sim10^{44-45}\Qunit$ at all particle energies}
\author{Boaz Katz\altaffilmark{1}, Eli Waxman\altaffilmark{2}, Todd Thompson\altaffilmark{3,4}, Abraham Loeb\altaffilmark{5}}
\altaffiltext{1}{Institute for Advanced Study, Einstein Drive, Princeton, New Jersey, 08540, USA}
\altaffiltext{2}{Particle Physics \& Astrophysics Dept., Weizmann Institute of Science, Rehovot 76100, Israel}
\altaffiltext{3} {Department of Astronomy, The Ohio State University, 140 W. 18th Ave., Columbus, OH, 43210, USA}
\altaffiltext{4} {Center for Cosmology \& Astro-Particle Physics, The Ohio State University, 191 West Woodruff Ave., Columbus, OH, 43210, USA}
\altaffiltext{5}{Harvard-Smithsonian Center for Astrophysics, Cambridge, MA 02138, USA}

\begin{abstract}
The energy output (per logarithmic interval of particle energies) of Cosmic Rays (CRs) with energies $10\GeV\lesssim\Ep\lesssim100\GeV$ is $\sim 10^{47}\erg$ per solar mass of star$-$formation, based on the CR production rate in the Milky Way and in starburst galaxies, implying a generation rate of $\Ep^2 \QUp\sim 10^{45}\Qunit$ in the local universe. It is only $\sim 10$ times larger than the output,
$\Ep^2 \QUp=0.5\pm0.2\X10^{44}\Qunit$, of Ultra High Energy CRs (UHECRs) at energies $10^{10.5}\GeV<\Ep<10^{12}\GeV$ (obtained assuming they are mostly protons), which in turn is comparable to the lower limit of $\Ep^2 \QUp\ge 0.5\X10^{44}\Qunit$ of high energy CRs with $10^6\GeV\lesssim\Ep\lesssim 10^{8}\GeV$ implied by the saturation of the Waxman-Bahcall bound by the neutrino excess recently discovered by IceCube. These
similarities are consistent with a flat production spectrum, $\Ep^2 \QUp\sim \text{const}$ for CRs at all observed energies.
If a flat production spectrum is generated by our galaxy, the observed CR flux in the range $10^{6.5}-10^{9.5}\GeV$,  above the "knee",  is suppressed compared to lower energies due to propagation effects rather than acceleration upper limits. As suggested by Parizot and Aublin, the most exciting possibility is that cosmic rays at all energies are emitted from a single type of (unknown) sources, which can not be supernova remnants.
\end{abstract}

\section{Introduction}
The origin of the observed Cosmic Rays (CRs) at different energies is still unknown \citep[see][for
reviews]{Blandford87,Axford94,Nagano00,Helder2012,Lemoine2013}. The energy density (per logarithmic particle energy) changes by about 8 orders of magnitudes across the observed particle energy range of $10^{9}\eV-10^{20}\eV$.  The cosmic ray spectrum steepens around $\sim 5\times 10^{15}$~eV (the ``knee'') and flattens around $5\times 10^{18}$~eV
(the ``ankle'').  Below the knee the cosmic rays are thought to originate from Galactic supernovae. Above the ankle, the so called Ultra High Energy CRs (UHECRs) are believed to be of extra-Galactic (XG) origin since they cannont be confined by the galactic magnetic field and their measured flux is nearly isotropic.

In this letter we use current data to estimate the production rate of CRs in the local universe at the different energies and show that CRs of all observed energies may be produced with a universal flat energy production spectrum $\Ep^2 d\dot n/d\Ep\sim \text{const}$. A softer universal production spectrum was suggested by \citet{Parizot05,Aublin05} with a production rate of low energy CRs which is 1000 times larger than UHECRs. We resolve the differences and show that the softer spectrum found by these authors is due to over(under)estimates of the energy production rates at low(high) energy.

\section{Energy production rates at low, intermediate and Ultra-High energies}\label{sec:Energetics}
If CRs at all energies are produced in galaxies, the observed flux of CRs at low energies is enhanced by several orders of magnitude due to the confinement of these CRs in our galaxy \citep{Loeb02} and the CR production spectrum is much harder than the observed spectrum. We next provide estimates for the production of CRs at the different energies for which we have reliable constraints. The estimates are summarized in figure 1.

\begin{figure}
\includegraphics[scale=0.5]{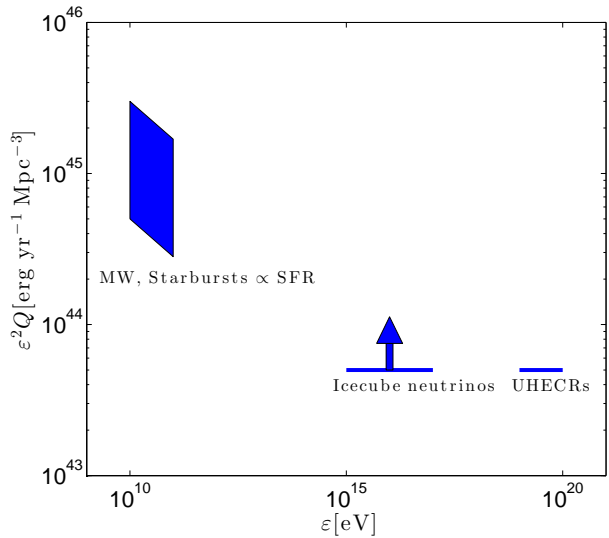}
\caption{Constraints on the energy production rate density of cosmic-rays (CRs) in the local universe (per logarithmic unit of particle energy). The production of CRs with $\vep\sim 10^{10-11}\eV$ (shaded area) is estimated in Eqs. \eqref{eq:LE} and \eqref{eq:RateLow} based on the production in our galaxy and in starbursts galaxies, assuming it follows the star formation rate (SFR). The lower bound on CRs with  $\vep\sim 10^{15-17}\eV$ is obtained  using Eq. \eqref{eq:WB} from the neutrino flux detected by Icecube which is assumed to be extra galactic. The production of Ultra-high-energy CRs (UHECRs) with  $\vep\sim 10^{19-20}\eV$ (solid line) is based on the observed flux of these CRs, assuming they are mainly protons and taking into account the interactions with the Cosmic-Microwave-Background (CMB) and is given in Eq. \eqref{eq:dPCRdV19}.
\label{fig:production}}
\end{figure}

\subsection{
UHE energies, $\Ep\sim 10^{19}-10^{21}\eV$}\label{sec:UHECRs}
Consider first the energy production of the highest energy cosmic rays, $\Ep\sim 10^{19}-10^{21}\eV$. The distance of the extra-galactic sources of the observed CRs in this range is limited by propagation losses due to the interaction with the inter-galactic radiation field. The composition is controversial, with air-shower data from the Fly's Eye, HiRes and Telescope Array observatories suggesting a proton dominated composition \citep[][]{Bird1993,Abbasi10,Sagawa2011} while the Pierre Auger Observatory suggesting a transition to heavy elements above $10^{19}\eV$ \citep[][]{Abraham10}. Due to this discrepancy, and due to the experimental and theoretical uncertainties in the relevant high energy particle interaction cross sections used for modeling the shape of the air showers, it is impossible to draw a definite conclusion regarding composition based on air-shower data at this time. It should be noted that the anisotropy signal measured at high energies, combined with the absence of this signal at low energies, is an indication for a proton dominated composition at the highest energy \citep{LemoineWaxman09,Liu13}. However, the anisotropy signal is so far identified with only a $\sim2\sigma$ confidence level \citep[see e.g.][for recent discussions]{Waxman11Review,Lemoine2013}. Below we assume that protons have a significant contribution to the flux and provide a quantitative estimate for their production under this assumption.

The energy production rate of UHECRs in the range $10^{19.2}\eV<\Ep<10^{20}\eV$ is \citep{Waxman95b,Bahcall03,Katz09}
\begin{equation}\label{eq:dPCRdV19}
\Ep^2\QUp(30 {\rm EeV}) \sim \Ep^2\frac{dn}{d\Ep} /t_{\rm eff}\sim 0.5\times10^{44}\Qunit
\end{equation}
where $\Ep^2dn/d\Ep$ is the energy density of observed CRs per logarithmic particle energy and $t_{\rm eff}$ is the effective time that the UHECRs propagate before losing energy. In the range $10^{19.2}\eV<\Ep<10^{19.6}\eV$, the effective time is approximately constant and is well approximated by \citep{Katz09}
\begin{equation}\label{eq:t_eff}
t_{\rm eff}\approx(\alpha-1)(\tau_{0,ep}^{-1}+2H_0)^{-1}\approx 2.5\times10^{9}\yr,
\end{equation}
where $H_0\approx (14\times 10^9\yr)^{-1}$ is the Hubble constant, $\tau_{0,ep}^{-1}= \dot\Ep/\Ep\approx (4\times 10^9\yr)^{-1}$ is the energy loss rate of CRs due to pair production interactions with the Cosmic Microwave Background (CMB) and $\alpha\approx 2$ is the CR generation spectral index $\QUp\propto \Ep^{-\alpha}$. The observed flux at $\Ep=10^{19.5}\eV$ is roughly $\Ep^2dn/d\Ep\approx10^{-20.3}\erg\cm^{-3}$ \citep[e.g.][and references therein]{Katz09}, leading to Eq. \eqref{eq:dPCRdV19}. Beyond $10^{19.6}\eV$, the effective time drops quickly with energy due to pion production \citep{Greisen66,Zatsepin66}. This predicted Greisen-Zatsepin-Kuzmin (GZK) cutoff is clearly observed in the flux \citep[e.g.][]{Bahcall03,Abbasi08,Abraham08} and the inferred generation rate is roughly constant to at least $10^{20}\eV$\citep[e.g.][]{Bahcall03, Katz09}, beyond which the flux is poorly measured.

The rate estimate, Eq. \eqref{eq:dPCRdV19}, is based on the direct measurement of UHECRs and the well understood physics of proton-CMB interaction and is accurate (to $\sim 30\%$) as long as the composition is dominated by protons. If the composition is dominated by heavier nuclei (up to iron), the energy generation rate at $\Ep\sim 10^{19.5}\rm eV$ may change by a factor of $\sim$ few \citep[given the similarity of the attenuation lengths of nuclei and protons, see e.g. figure 2 in ][]{Allard12}.

\subsection{
Intermediate energies, $10^{15}{\rm eV}\lesssim \Ep\lesssim 10^{18}\rm eV$}\label{sec:Intermediate}
The recent detection of high energy neutrinos by the IceCube collaboration \citep{IceCubeDetection,Laha13}, implies a lower limit on the energy production of the CRs which produce these neutrinos. We next briefly describe this constraint. For a more detailed discussion, see \citep{Spector13}.

The IceCube collaboration has recently reported the detection of 26 neutrinos in the energy range of 50~TeV to 2~PeV, which constitutes a $4\sigma$ excess above the expected atmospheric neutrino and muon backgrounds \citep{IceCubeDetection}. The excess neutrino spectrum is consistent with $dn/d\Enu \propto \Enu^{-2}$, its angular distribution is consistent with isotropy, and its flavor ratio is consistent with $\nu_e:\nu_\mu:\nu_\tau=1:1:1$. We stress that the spectral shape, angular distribution and composition are currently poorly constrained due to the low statistics. The best fit normalization of the intensity is $\Enu^2\Phi_\nu=3.6\pm1.2\times10^{-8}{\rm GeV/cm^2 s\, sr}$, coinciding (in normalization and spectrum) with the Waxman-Bahcall (WB) bound on the neutrino intensity that may be produced by extra-Galactic sources \citep{WBbound},
\begin{align}
\Enu^2\Phi_{\rm WB,\, all\, flavor}&=3.4\times10^{-8}{\rm \frac{GeV}{cm^2s\,sr}}\cr
&\times \frac{\xi_z}{3}\frac{(\Ep^2\QUp)_{z=0}}{0.5\times10^{44}\Qunit}
\label{eq:WB}
\end{align}
where $\Ep^2\QUp$ is the UHECR proton production rate, and $\xi_z$ is (a dimensionless parameter) of order unity, which depends on the redshift evolution of $\Ep^2\QUp$. The value $\xi_z=3$ is obtained for redshift evolution following that of the star-formation rate or AGN luminosity density, $\Phi(z)=(1+z)^3$ up to $z=2$ and constant at higher $z$ ($\xi_z=0.6$ for no evolution).

The neutrino excess cannot originate from interaction of cosmic-ray protons with interstellar gas in the Galaxy, which produces an average (over angles) intensity of $\approx10^{-9}(\vep/100{\rm TeV})^{-0.7}{\rm GeV/cm^2 s\, sr}$ (based on the Fermi determination of the $\pi_0$ decay intensity at 100~GeV \citep{Ackermann12} and the Galactic CR  spectrum $\vep^2dn/d\vep\propto \vep^{-0.7}$). It is also  unlikely to be due to (unknown) Galactic sources, which are expected to be strongly concentrated along the galactic disk. The coincidence with the WB flux also suggests an extra-Galactic origin. Note, that a $\nu_e:\nu_\mu:\nu_\tau=1:1:1$ flavor ratio is consistent with that expected for neutrinos originating from pion decay in cosmologically distant sources, for which oscillations modify the original $1:2:0$ ratio to a $1:1:1$ ratio \citep[][]{Learned95,Athar06}.

The IceCube excess neutrinos are likely produced by interactions of high energy CR protons with protons or photons, or of high energy CR nuclei with protons, which produce pions that decay to produce neutrinos. Assuming that the neutrino excess is due to extra-Galactic sources, a lower limit of $\Ep^2\QUp\ge0.5\times10^{44}\Qunit$ on the local, $z=0$, CR production rate is implied. This is due to the fact that the upper bound is obtained by assuming that CR protons are produced at this rate and lose all their energy to pion production. Note, that the limit obtained for nuclei is similar, since photo-disintegration does not reduce significantly the energy per nucleon. The CR energy range corresponding to the energy range of the observed neutrinos is $\approx 1A-100A$~PeV, where $A$ is the atomic number of the CRs \citep{Spector13}. We note that the observed cutoff in the neutrino spectrum above a few PeV \citep{IceCubeDetection, Laha13} may be due to efficient escape of the CRs from the environments in which they produce the pions and need not imply a cutoff in the CR production spectrum \citep[as suggested, e.g., by][for sources residing in starburst galaxies]{LWstarbursts}.

The coincidence of the observed neutrino flux and spectrum with the WB bound flux and spectrum is unlikely a chance coincidence. The observed neutrinos may be produced by sources with $\Ep^2\QUp\sim0.5\times10^{44}\Qunit$ and in which CRs lose most of their energy to pion production, or by sources with $\Ep^2\QUp\gg0.5\times10^{44}\Qunit$ and in which CRs lose only a small fraction of their energy to pion production. In the latter case, the small energy loss fraction should compensate the large energy production rate to reproduce the observed flux and spectrum. We thus consider the former option, sources with $\Ep^2\QUp\sim0.5\times10^{44}\Qunit$ and in which CRs lose most of their energy to pion production, to be more likely.

\subsection{
Low energies, $10{\rm GeV}\lesssim\Ep\lesssim 100\rm GeV$}
\label{sec:Production}

We estimate the low energy CR production rate in two different ways. One based on the CR production rate in starburst galaxies (via both their radio and gamma-ray emission), and the other based on the CR production rate in the Milky Way (MW).  The starburst/MW CR production rates are converted to a total production rate per unit volume in the local universe by assuming that they are proportional to the star formation rates (see discussion in section \sref{sec:discussion}).  Although the estimate based on the MW is less robust, since it is based
on a single galaxy at a particular time and since it is difficult to reliably estimate the MW production rate, both estimates agree to within a factor of a few.

In starburst galaxies, protons of energy $<1$~PeV are expected to lose most of their energy to pion production before escaping the galaxy, and the secondary $e^\pm$  produced by charged pion decay emit synchrotron radiation in the radio band \citep{LWstarbursts,Todd_gamma,Lacki2011}. Due to the strong magnetic fields in starburst galaxies \citep{Todd_B}, $e^\pm$ of energy $\sim1$~GeV, produced by $\sim10$~GeV protons, lose their energy by emitting $\sim1$~GHz radiation \citep{LWstarbursts}. Thus, the radio emission of starburst galaxies provides an estimate of the CR proton production rate: $\Ep^2\QUp(10{\rm GeV})\approx 12\nu L_\nu(1{\rm GHz})/f_{\rm synch}$, where $1/4$ of the energy of the charged pions, which carry $2/3$ of the proton energy, is deposited in the electrons and positrons, $\nu_{\rm synch}\propto \Epar_e^2$, and $f_{\rm synch}\lesssim1$ is the fraction of the $e^\pm$ radiated in synchrotron \citep{LWstarbursts,Lacki2011}. The radio luminosity of starbursts is linearly (and tightly) correlated with the FIR ($40-120\mu{\rm m}$) luminosity, $\nu L_\nu(1.4{\rm GHz})=1.7\times10^{-6} L({\rm FIR})$ \citep{Yun}, suggesting that $f_{\rm synch}$ is of order unity \citep[e.g.][]{Todd_B}. This conclusion is further supported by the comparison of the synchrotron and the gamma-ray emission of M82 and NGC 253, which imply $f_{\rm synch}\sim0.2$ \citep[assuming that the observed gamma-ray flux is dominated by $\pi_0$ decays][]{Lacki2011,TL2013}.  The FIR luminosity is, in turn, related to the SFR by ${\rm SFR}=1.7\times10^{-10}[L(8-1000\mu{\rm m})/L_\odot] M_\odot/{\rm yr}$, or equivalently $L(8-1000\mu{\rm m})/{\rm SFR}=7\times 10^{50} \erg/M_{\odot}$, with $\sim50\%$ uncertainty \citep{Yun,Kennicutt}. Using $L(8-1000\mu{\rm m})/L({\rm FIR})\sim1.3$ \citep{Yun},  we find that
\begin{equation}\label{eq:CR_SFR}
    \Ep^2\QUp(10{\rm GeV})/{\rm SFR}\approx 4\times10^{46}\,\left(\frac{0.3}{f_{\rm synch}}\right) \,{\rm erg}/M_\odot.
\end{equation}

Equivalently, a similar value can be obtained for M82 and NGC 253 by directly comparing their gamma rays to their FIR luminosities. The ratio of observed gamma-ray to FIR luminosities of both galaxies  is $\nu L_{\nu}({\rm GeV})/L(8-1000\mu{\rm m}\approx 1.3e-5 $ \citep{Ackermann12b} leading to CR efficiencies of $\Ep^2\QUp(10{\rm GeV})/{\rm SFR}\approx 3\times10^{46}{\rm erg}/M_\odot$, where we assumed that $1/3$ of the energy carried by $10\rm GeV$ CRs is converted to $\sim \rm GeV$  $\gamma$-rays.  
 
Using the local SFR density, $\dot{\rho}_{\rm SFR}(z=0)\sim 0.015 M_\odot\,{\rm yr^{-1}\,\,Mpc^{-3}}$ \citep[e.g.][]{HopkinsBeacom06}, we thus find
\begin{equation}\label{eq:LE}
   \Ep^2\QUp(10{\rm GeV})\sim 0.6\times10^{45}\Qunit.
\end{equation}
Note that if CRs are assumed to be produced by SNe, then eq.~(\ref{eq:CR_SFR}) and the number of SNe per stellar mass produced, $\sim10^{-2}/M_\odot$, imply $\Ep^2d{N}_p/d\Ep\sim 6\times10^{48}{\rm erg}$ per SN.

We next estimate the production rate of the same CRs by relating it to the total estimated production in our galaxy. Normalizing to the SFR, the production rate of CRs with energies $\Ep\sim 10^{10}-10^{11}\eV$ can be estimated by
\begin{equation}\label{eq:CR_SFR_MW}
\Ep^2\QUp(10{\rm GeV})/{\rm SFR}\sim\frac{\Ep^2 \QMWp}{\rm SFR(MW)}\sim 2\times 10^{47}{\rm erg}/M_\odot
\end{equation}
or
\begin{equation}\label{eq:RateLow}
\Ep^2\QUp(10{\rm GeV})\sim 3\times 10^{45}(\Ep/10\GeV)^{-0.25}\rm erg~Mpc^{-3}~\yr^{-1}
\end{equation}
where ${\rm SFR}({\rm MW})\sim 2 M_{\odot}\yr^{-1}$ is the current star formation rate in the Milky Way \citep[e.g.][]{Chomiuk11} and $\Ep^2\QMWp\sim4\X10^{47}(\Ep/10\GeV)^{-0.25}\erg\yr^{-1}$ is the production rate (per log particle energy) of CRs in our galaxy.

The quoted production rate of our galaxy is estimated by \citep[e.g.][]{Berezinskii90}
\begin{align}\label{eq:WCRGal}
\Ep^2 \QMWp\approx &c\Ep^2dn/d\Ep M_{\rm{gas}}/\Xesc\cr
\sim&4\X10^{47}(\Ep/10\GeV)^{-0.25}\erg\yr^{-1}
\end{align}
where $M_{\rm{gas}}\sim 10^{10} M_{\odot}$ is the mass of the gas in the Galaxy
\citep[][and references therein]{Naab06},
\begin{equation}\label{eq:CRlow}
\Ep^2dn/d\Ep\approx 0.12(\Ep/10\GeV)^{-0.75}\eV\cm^{-3}
\end{equation}
is the proton flux at the solar neighborhood \citep[e.g.][]{Moskalenko02} and
\begin{equation}\label{eq:sigLowX}
\Xesc\approx 8.7 \left(\frac{\Ep/Z}{10 \GeV}\right)^{-0.5}\gr\cm^{-2},
\end{equation}
is the average grammage traversed by the CRs which is deduced from the abundance of spallation secondaries \citep[mainly from the ratio Boron/Carbon, e.g.][]{Engelman90,Jones01,Webber03}.

 Equation \eqref{eq:WCRGal} is insensitive to the poorly constrained values of the CR confinement volume and confinement times. It can be derived directly using the following argument: The integrated grammage of all CRs (in an infinitesimal particle energy interval $\Ep,\vep+d\Ep$) over a long time $T$ (longer than the escape time from the Galaxy) can be expressed in two ways: (i) The total amount of particles generated (within the particle energy interval), $T\QMWp$ times the average grammage each traversed $\Xesc$ (ii) The instantaneous total rate of grammage being traversed, $dn~M_{\rm{gas}}c$, times the total time $T$, where $dn$ is the density of CRs within the particle energy interval. By equating these expressions, $T\QMWp\Xesc=dn~M_{\rm{gas}}cT$, Eq. \eqref{eq:WCRGal} is obtained (by dividing by $d\Ep$). It is assumed however that the flux of CRs is roughly uniform throughout the galactic disk and that the average grammage is similar to that in the solar neighborhood. If CRs cannot propagate freely across the Galaxy, this estimate may have a large error. However, in this latter case, a local estimate would be more accurate and yields approximately the same value. Assuming a local gas surface density of $10M_{\odot}\rm pc^{-2}$ and local SFR of $2M_{\odot}\rm pc^{-2}\rm Gyr^{-1}$ \citep[e.g. ][]{Kennicutt12}, would result in the same generation rate per SFR as implied by Eq. \eqref{eq:RateLow}.

If, as currently widely believed, CRs originate from SNe, Eq. \eqref{eq:RateLow} would require an energy per supernova (per log particle energy) of $\Ep^2\QUp/\dot n_{\rm SN}\sim 2\times10^{49}\rm erg$ where $\dot n_{\rm SN}\sim 10^{-4}\rm Mpc^{-3}$ is the rate of supernovae at $z=0$ \citep{Horiuchi09}.

The fact that the CR production estimate based on the Milky Way and that based on star-burst galaxies deviate by $\sim 5$ should not be surprising given the large uncertainties in each of these estimates. In particular the star-formation estimates are uncertain by a factor of at least 2. We conclude that the current rate of CR production is of order $10^{45} \Qunit$.

\subsection{Comparison with earlier work}
In  \citep{Parizot05,Aublin05}, the production rate of CRs was estimated to be 1000 times higher than that of UHECRs using the same basic assumption that other galaxies have acceleration efficiencies which are similar to ours. This ratio is much higher than the ratio of $\sim 10$ we estimated above. After carefully comparing our analysis we found that the main difference is due to the fact that \citet{Parizot05,Aublin05} ignored the interactions of the UHECRs with the CMB and therefore underestimated their production rate. For example, in \citep{Aublin05}, the effective life time of UHECRs was estimated to be $25H_0^{-1}e^{-0.8\alpha}\approx 6\times 10^{10}\rm yr$ which is about $20$ times longer than the actual time they can propagate before losing a significant amount of energy due to the interaction with the CMB and adiabatic losses [see Eq. \eqref{eq:t_eff}]. In addition, the halo size and escape time adopted by these authors for estimating the galactic CR production rate is not compatible with the measured grammage (a factor of $\sim 4$ discrepancy). This last inaccuracy was partially canceled by their use of an effective galaxy density of $n_{\rm gal}\sim 3\times 10^{-3}\rm Mpc^{-3}$ rather than our scaling by SFR which is equivalent to an effective galaxy density of ${\rm SFR}(z=0)/{\rm SFR(MW)}\sim 7.5\times10^{-3}\rm Mpc^{-3}$.

%
%

%

\section{Discussion}\label{sec:discussion}
In this letter we showed that the production of low energy CRs ($\Ep \sim10^{10}\eV$) in the local universe is $\vep^2_pQ\sim 10^{45}\Qunit$ (Eqs. \eqref{eq:LE} and \eqref{eq:RateLow}, with a large uncertainty) and is not much higher ($\sim 10$ times higher) than the lower limit for production of extra-galactic CRs with intermediate energies ($\Ep \sim 10^{17}\eV$) and ultra-high energies  ($\Ep \gtrsim 10^{19.5}\eV$). It is thus possible that all CRs are generated in galaxies like our own with an energy production efficiency which is very slowly declining across the 10 orders of magnitude of CR energies (see figure 1).

The similarity between the energy production rates of CRs throughout the particle energy range  $10^{10} \eV$ to $10^{19} \eV$, is consistent with a universal CR energy production rate in
galaxies with a CR spectrum of $\Ep^2dn/d\Ep\propto \Ep^{-\alpha}$ with $\alpha\approx 2$. This power-law slope is expected for
astrophysical sources which accelerate particles in strong collisionless shocks, including non relativistic
\citep[e.g.][]{Krymskii77, Axford77, Bell78, Blandford78} and relativistic
\citep[e.g.][]{Kirk87,Bednarz98,Achterberg01,Keshet05,Keshet06,Katz07,Spitkovsky08} shocks, and is found in supernovae
\citep[e.g.][]{Reynolds92} and $\gamma$-ray bursts \citep[e.g.][]{Waxman97a}.

If CRs are produced with the suggested spectrum in galaxies like our own, including our own, the generation spectrum at high energy, $\Ep \gtrsim 10^{17}\eV$, within our galaxy should be
\begin{equation}
\Ep^2 \QMWp\sim 10^{46}\rm erg~yr^{-1}
\end{equation}
and would imply an averaged local energy density of at least $\Ep^2 n\gtrsim \Ep^2 \QMWp/(A_{\rm MW}c)\sim 2\times 10^{-17}\rm ergs~cm^{-3}$, where $A_{\rm MW}\sim 1000\rm kpc^2$ is the area of the Galaxy. This energy density is much higher than the observed energy density above the knee. This apparent discrepancy can be reconciled only if the sources are transient \citep{Loeb02}, and the flux we are currently observing is low compared to the averaged flux.

While the (very rough) apparent similarity of production rates at low, intermediate and ultra-high energies may well be a coincidence, the most exciting possibility is that there is one type of such transient sources that emits the CRs at all observed energies \citep{Parizot05,Aublin05}.
In this case, the rate of occurrence of these sources and their energy output is constrained by the observed CR spectrum. Their energy output should explain that of the Milky Way implying $\dot N_{\rm MW}E_{\rm CR}\sim 10^{48}\rm erg \yr^{-1}$ [see Eq. \eqref{eq:WCRGal}], while their rate should be lower than the CR confinement region's light crossing time to allow the dim state observed at the energies above the knee \citep{Loeb02}. The scale height of the CR confinement region at very high energies is unknown and can be anywhere between the disk scale height $\sim 300 \rm pc$ and the size of the Galaxy $\sim 10{\rm kpc}$. These two limits imply very different (maximal) event rates of $(1000 \rm yr)^{-1}$ and $(3\times 10^4 \rm yr)^{-1}$ in the local $300 \rm pc$ region or the entire galaxy respectively. These rates imply in turn a wide range of allowable values for the CR energy release per event of $E_{\rm CR}\gtrsim 10^{47}\rm erg$ and $E_{\rm CR}\gtrsim 3\times 10^{52}\rm erg$. The range of allowable energy outputs may be reduced by reducing the uncertainty in the CR disk scale height.

\acknowledgments We thank J. F. Beacom and B. Lacki for discussions leading to this work and for useful comments.
B. K. is a John Bahcall fellow. EW is partially supported by ISF and IAEC-UPBC grants.

\bibliographystyle{apj}

\end{document}

%% file: Definitions.tex
\newcommand{\DDir}{\relax{D\kern-.7em{/}}}

\newcommand{\X}{\times}

\newcommand{\be}{\begin{equation}}
\newcommand{\ee}{\end{equation}}
\newcommand{\bea}{\begin{equation*}}
\newcommand{\eea}{\end{equation*}}

\newcommand{\nin}{\relax{\in\kern-.8em{/}}}

\newcommand{\vep}{\varepsilon}

\newcommand{\cm}{\mbox{ cm}}

\newcommand{\yr}{\mbox{ yr}}

\newcommand{\erg}{\mbox{ erg}}

\newcommand{\Mpc}{\mbox{ Mpc}}
\newcommand{\eV}{\mbox{eV}}

\newcommand{\GeV}{\mbox{~GeV}}

\newcommand{\gr}{\mbox{g}}

\newcommand{\sref}{\S~\ref}